\documentclass[12pt]{article}

\usepackage{Adam}
\usepackage{authblk}
\usepackage[cm]{fullpage}
\usepackage{cite} 

\newcommand{\tildeQ}{\tilde Q}
\renewcommand{\P}{P}
\newcommand{\Q}{Q}

\newcommand{\Ep}{E_P}
\newcommand{\tfin}{T}
\newcommand{\inttf}{\int_0^\tfin}
\newcommand{\dtau}{\dd\tau}
\newcommand{\dt}{\dd t}

\newcommand{\nphysical}{{n_s}}
\newcommand{\nlogical}{n}
\newcommand{\nphysperlog}{\ell}

\newcommand{\Hsys}{H_{\text{comp}}^{\text{L}}}
\newcommand{\Henv}{H_{\text{env}}}

\newcommand{\physdummy}{s} 
\newcommand{\paulidummy}{\mu}

\newcommand{\ddt}[1][]{\frac{\dd #1}{\dd t}}
\newcommand{\bigO}[1]{\mathcal{O}\!\left(#1\right)}

\newcommand{\bignorm}[1]{\Big\Vert #1 \Big\Vert}

\newcommand{\eqn}[1]{Eq.~\eqref{#1}}
\newcommand{\eqns}[1]{Eqs.~\eqref{#1}}

\newcommand{\placefidplot}[1]{{\includegraphics[width=18.2cm, height=7.8cm]{#1}}}
\newcommand{\placeprobplot}[1]{{\includegraphics[width=14cm, height=6cm]{#1}}}

\title{\textbf{Error suppression in Hamiltonian based quantum computation using energy penalties}}

\date{July 4, 2014}

\author{
Adam D. Bookatz\thanks{bookatz@mit.edu},\quad
Edward Farhi\thanks{farhi@mit.edu},\quad
Leo Zhou\thanks{leozhou@mit.edu}
}

\affil{Center for Theoretical Physics, Massachusetts Institute of Technology, Cambridge, Massachusetts 02139, USA}

\begin{document}
\maketitle

%
%
\abstract{
We consider the use of quantum error detecting codes, together with energy penalties against leaving the codespace, as a method for suppressing environmentally induced errors in Hamiltonian based quantum computation. 
This method was introduced in \cite{JFS06} in the context of quantum adiabatic computation, but we consider it more generally. 
Specifically, we consider a computational Hamiltonian, which has been encoded using the logical qubits of a single-qubit error detecting code, coupled to an environment of qubits by interaction terms that act one-locally on the system. Energy penalty terms are added that penalize states outside of the codespace.
We prove that in the limit of infinitely large penalties, one-local errors are completely suppressed, and we derive some bounds for the finite penalty case. Our proof technique involves exact integration of the Schrodinger equation, making no use of master equations or their assumptions.
We perform long time numerical simulations on a small (one logical qubit) computational system coupled to an environment and the results suggest that the energy penalty method achieves even greater protection than our bounds indicate.
}

\qquad\qquad\quad\ \, {{\sc mit-ctp} 4564}

%
\section{Introduction}
%

A major problem on the road to building scalable quantum computers is the difficult task of protecting the system from errors, such as those due to unwanted environmental interactions. In the usual circuit model of quantum computation, the theory of quantum error correction has been well-developed, culminating in the threshold theorem \cite{abo97, kit97c, kit97b, got97, pre98c}, which proves that, provided the error rate in a quantum computing system can be reduced to below a certain threshold, errors can be suppressed arbitrarily well using quantum error correcting codes. 
The situation for the Hamiltonian model of quantum computation as used in, for example, adiabatic quantum computing, continuous-time quantum walks, and Hamiltonian simulation problems, is less understood and no fault-tolerant theorem is known. In this paper, we take steps towards establishing such a theorem.

In the Hamiltonian model, the computational system is described by a Hamiltonian, which is a (possibly time-dependent) Hermitian operator, $H_\text{comp}$, that governs the time-evolution of the system according to
\[
	i\ddt\ket{\phi(t)} = H_\text{comp}(t)\ket{\phi(t)} \,,
\]
where $\ket{\phi(t)}$ is the state of the computational system at time $t$. In this model, the goal is to evolve some initial state $\ket{\phi(0)}$ to a final state $\ket{\phi(\tfin)}$, the measurement of which reveals some information about the problem to be solved. Note that no instantaneous unitary gates are applied, nor are any intermediate measurements performed.
To consider the effects of unwanted environmental interaction, one must consider the Hamiltonian $H_\text{comp} + H_\text{environment} + H_\text{interaction}$ that governs the evolution of the entire system-environment supersystem. The goal of error suppression is to ensure that the state of the system at time $\tfin$ is approximately as though the evolution had been governed by just $H_\text{comp}$ alone.

It is not clear how to adapt the successful error correcting code techniques of the circuit model to the Hamiltonian model. In a conventional quantum error correcting code \cite{NCtext}, each qubit is encoded as a logical qubit, comprised of several physical qubits, so that the occurrence of any single-qubit error on any physical qubit can be detected.
The use of such a code in the error correcting circuit model essentially consists of four steps: the state is encoded, the state is allowed to evolve, a measurement is made to determine what error has occurred (if any), and gates are applied to correct that error.
In our Hamiltonian model, we do not allow intermediate measurements or the application of instantaneous gates, and therefore rule out any active determination and correction of errors; thus, a different strategy is required.

The error suppression strategy used in this paper is that of \textit{energy penalties}, first suggested in \cite{JFS06}, in which the system Hamiltonian is modified according to a quantum error detecting code and a constant (time-independent) term is added to the Hamiltonian. This extra term, the energy penalty, penalizes states that have been corrupted by, say, single-qubit errors. It is believed that such a penalty will suppress the occurrence of environmentally induced errors, as it imposes an energy barrier that must be surmounted for an error to occur. In this work, we prove that, in principle, this energy penalty method does indeed work; we show that it successfully suppresses errors arbitrarily well when the penalty is arbitrarily large. (Throughout the paper we concentrate on 1-local errors and use a 1-qubit error detecting code. In the appendix, however, we show that this result can be generalized to $k$-local errors when using a $k$-qubit error detecting code.) We also explore (in the 1-local error case) how well the penalty terms work when the penalty is not infinite but of a reasonable size. We then show the results of small-system numerical simulations that suggest that the achieved protection is even better than our bounds can predict.

We note that since we will not be performing active error correction, we do not need an error \textit{correcting} code, which gives information about which error occurred; rather, it suffices to use an error \textit{detecting} code, which only detects whether any error has occurred.

The energy penalty method remains true to the Hamiltonian model paradigm, in that the evolution is still specified by a Hamiltonian. Other previously suggested Hamiltonian model error suppression methods, for example exploiting the Zeno effect \cite{MS77, PRDL12} or using dynamical decoupling \cite{VKL99,Lidar08}, usually require abilities outside of the Hamiltonian paradigm: in the Zeno case, intermediate measurements are required, and in the dynamical decoupling case, instantaneous unitary pulses are required. While some dynamical decoupling based methods, such as Eulerian dynamical decoupling \cite{VK02} and its extension to Hamiltonian simulation \cite{BWV14}, do suppress environmental interactions while ensuring desired system evolution by adding control terms to the Hamiltonian, those terms are rapidly time-dependent. In contrast, the method of energy penalties used in this paper is passive, since it is achieved by adding a constant term to the encoded Hamiltonian, and would be useful even when fast, active control is not available. A discussion of the similarity between the energy penalty, Zeno, and dynamical decoupling methods can be found in \cite{YSB13, FLP04}.

%
\section{Quantum error detecting codes}
%
We first review some basic facts about quantum error detecting codes.
Suppose that we have a $[[\nphysperlog,1]]$ quantum error detecting code, meaning that by encoding a single qubit as a logical qubit comprised of $\nphysperlog$ physical qubits, we can detect arbitrary 1-qubit errors. Throughout this paper, we use this code to protect our system of $\nlogical$ qubits, meaning that each qubit of the original $H_\text{comp}$ is encoded to be $\nphysperlog$ qubits, so that the full encoded system consists of $\nphysical = \nphysperlog\nlogical$ qubits.

Specifically, for each qubit register $i$, the original computational basis states $\ket{0}_i$ and $\ket{1}_i$ are encoded as the $\nphysperlog$-qubit logical states $\ket{0_L}_i$ and $\ket{1_L}_i$. The codespace of the $i$th logical qubit is then the span of the logical states, $\lbrace a\ket{0_L}_i+b\ket{1_L}_i : |a|^2+|b|^2=1 \rbrace$.
Associated with this codespace is the projection operator 
$$
	\P_i = \selfketbra{0_L}_i + \selfketbra{1_L}_i \,,
$$ 
where $\P_i$ acts as the identity on all physical qubits other than those associated with the logical qubit $i$.
Note that states in the codespace are invariant under $\P_i$, whereas $\P_i$ kills states that are orthogonal to the codespace of the $i$th qubit.

Saying that the code can detect arbitrary 1-qubit errors is equivalent to saying that the code detects all single-qubit Pauli errors, i.e. an error caused by the application of a Pauli operator ($X$, $Y$, or $Z$) to any single physical qubit. Thus, for any single Pauli operator $\sigma$ acting on one of the $\nphysperlog$ physical qubits comprising  logical qubit $i$, we have
\begin{equation}\label{eq:PsigmaP}
	\P_i \sigma \P_i = 0 \,.
\end{equation}
The full codespace for the entire logical space (over all $\nlogical$ logical qubits) corresponds to the projector 
\begin{equation}\label{eq:Pprod}
	\P = \P_1 \P_2 \cdots \P_\nlogical \,.
\end{equation}

The quantum code also allows us to "encode" the Pauli operators $X$, $Y$, and $Z$ as \textit{logical operators} $X_L$, $Y_L$, and $Z_L$. Logical operators are Hermitian operators that have the same effect on the logical basis states as their corresponding Pauli operators have on the corresponding basis states.
Furthermore, the logical operators associated with qubit $i$ commute with the codespace projector $\P_i$, i.e. $X_L \P_i = \P_i X_L $, and similarly for $Y_L$ and $Z_L$.

As a concrete example, consider the 4-qubit code of Jordan-Farhi-Shor\cite{JFS06}, in which
\[
\begin{split}
	\ket{0_L} &= \half\Big(\ket{0000} + i\ket{0011} + i\ket{1100} + \ket{1111}\Big) \\
	\ket{1_L} &= \half\Big(-\ket{1010} + i\ket{1001} + i\ket{0110} - \ket{0101}\Big) \\
	X_L &= ~~ Y\otimes\id \otimes Y\otimes\id \\
	Y_L &= -\id\otimes X\otimes X \otimes\id \\
	Z_L &= ~~ Z\otimes Z\otimes\id\otimes\id \,.
\end{split}
\]
Observe that the logical operators have the same effect on logical qubits as do the operators to which they correspond have on unencoded qubits; e.g. $X_L\ket{0_L} = \ket{1_L}$. 

Using the logical operators, we can encode the Hamiltonian that acts on the system. Suppose that $H_\text{comp}$ is some Hermitian operator on the original ($\nlogical$-qubit) system. Because the Pauli matrices (along with the identity) form a basis for all $2\times 2$ matrices, we may generically write 
$$
	H_\text{comp}(t) = \sum_{\substack{\sigma_i\in\lbrace\id,X,Y,Z\rbrace \\ i=1,\ldots,\nlogical}} c_{\sigma_1,\ldots,\sigma_\nlogical}(t) \  \sigma_1\otimes\cdots\otimes\sigma_\nlogical ,
$$ 
where the sum is over all possible choices of $\sigma_i\in\lbrace\id,X,Y,Z\rbrace$ for each $i$. We may therefore encode the Hamiltonian by replacing $X,Y,Z$ with $X_L,Y_L,Z_L$ in the sum above, to obtain
$$
	\Hsys(t) = \sum_{\substack{\sigma_i\in\lbrace\id,X_L,Y_L,Z_L\rbrace \\ i=1,\ldots,\nlogical}} c_{\sigma_1,\ldots,\sigma_\nlogical}(t) \  \sigma_1\otimes\cdots\otimes\sigma_\nlogical ,
$$ 
which is a Hamiltonian on the $\nphysical$-qubit encoded space built entirely out of logical operators (and $\id$). Since each logical operator commutes with each $\P_i$, $\Hsys$ also commutes with each $\P_i$ and with $\P$.

Observe that the logical operators in the Jordan-Farhi-Shor code are all 2-local. The encoding in this case thus doubles the locality of the original Hamiltonian, so that if the original Hamiltonian is 2-local, the encoded one is 4-local. As \cite{JFS06} points out, such an encoding is optimal (in terms of locality) if the code is to protect against arbitrary 1-qubit errors.

%
\section{The Hamiltonian model and energy penalties}
%
In this paper we consider a system coupled to an environment. We do not attempt to modify the environment or the system-environment interaction. However, we assume that we can modify the Hamiltonian of the system, and do so in two ways. As just discussed, we encode the original computational Hamiltonian in a quantum code. Furthermore, we add extra terms (acting only on the system) that penalize system states that are outside of the codespace. 

The combined system-environment Hamiltonian, $H$, after encoding and penalty modifications, consists of three parts, and can be written as
\begin{equation*}
	H = H_0 + \lambda V + \Ep \tildeQ \,.
\end{equation*}
We discuss each of these parts in turn.

\begin{enumerate}
\item 
The first term is 
$$
	H_0 = \Hsys\otimes\id_{\text{env}} + \id_{\text{sys}}\otimes \Henv \,,
$$
which governs the evolution in the absence of any system-environment interaction. Both $\Hsys$ and $\Henv$ are in general time-dependent.
Evolution under $\Hsys$ alone is equivalent to evolution under $H_\text{comp}$ and represents the desired evolution we wish to protect.

Because the system Hamiltonian is encoded, the system consists of $\nphysical=\nphysperlog\nlogical$ qubits. The size of the environment will play no role in our discussion, except when we do simulations, and can be thought of as much larger than the system size. \\
Note that $\Hsys$ is built up from only logical operators and therefore commutes with each $\P_i$. Since $\Henv$ (which acts only on the environment) trivially commutes with each $\P_i$, we have
\[
	[H_0,\P_i]=0 \text{\quad for } i=1,\ldots,\nlogical .
\]

\item
$\lambda V$ is the error Hamiltonian, reflecting the coupling of the system to the environment, with $\lambda$ serving as a time-independent (presumably small) parameter indicating the strength of the interaction (with units of energy), and $V$ a  Hermitian operator acting on the full system-environment space. 
Our code is designed to protect against 1-qubit errors, so we assume a 1-qubit error model, i.e. that $V$ acts 1-locally on the system. Thus, we can write $V$ as a sum of terms
\begin{equation}\label{eq:Vdefn}
	V = \sum_{\physdummy=1}^{\nphysical} \sum_{\paulidummy=X,Y,Z} \sigma_\paulidummy^\physdummy \otimes B_\paulidummy^\physdummy
\end{equation}
where $\sigma_\paulidummy^\physdummy$ is the $\paulidummy^{\text{th}}$ Pauli matrix acting on physical system qubit $\physdummy$ and each $B_\paulidummy^\physdummy$ is some operator acting on a small set of environmental qubits.
We also allow the possibility that $B_\paulidummy^\physdummy = \id$, which could represent 1-local system control errors independent of the environment.

For convenience, we group terms in $V$ according to the logical system qubit on which they act, so that
\begin{equation} \label{eq:Vsum}
	V = \sum_{i=1}^{\nlogical} V_i
\end{equation}
where each $V_i$ is an operator whose 1-local action on the system is only on the $\nphysperlog$ system qubits that comprise logical qubit $i$.
Observe that $V_i$ causes 1-local errors on the system, as per \eqn{eq:Vdefn}, and that we are using a code that can detect arbitrary 1-qubit errors, as per \eqn{eq:PsigmaP}. Thus, we have that 
\[
	\P_i V_i \P_i=0 \quad \text{for $i=1,\ldots,\nlogical$}
\]
which is crucial to our later analysis.

\item
$\Ep\tildeQ$ is our time-independent energy penalty, which penalizes states outside of the codespace. Specifically, $\Ep$ is a real constant with units of energy and $\tildeQ$ is the sum of the projectors $Q_i = \id-\P_i$, i.e.
\begin{equation}\label{eq:tildeQ}
	\tildeQ = \sum_{i=1}^{\nlogical} \Q_i = \sum_{i=1}^{\nlogical} (\id-\P_i) \,,
\end{equation}
so we have a separate energy penalty for each logical qubit.
In this context, $\tildeQ$ is to be understood as $\tildeQ\otimes\id_{\text{env}}$, since only the system is encoded. Observe that a state $\ketpsi$ is in the codespace if and only if $\tildeQ\ketpsi = 0$, so $\Ep\tildeQ$ applies an energy penalty of magnitude at least $|\Ep|$ to states outside of (i.e. orthogonal to) the codespace. 

We point out that $\tildeQ$ is the sum of codespace projectors, differing from the penalty used in \cite{JFS06} which is a sum of codespace stabilizer generators.
Note that the locality of $\tildeQ$ is that of each $\P_i$, which is at most $\nphysperlog$ (i.e. 4 in the case of the Jordan-Farhi-Shor code).

\end{enumerate}

The key point in this model is that $V$ acts precisely 1-locally on the system and we are using a quantum code that can detect 1-qubit system errors. This enables us to penalize the states that arise from the action of $V$, and therefore have hope of suppressing $V$'s effect. 
We can similarly consider the case in which $V$ acts $k$-locally on the system as long as the quantum code can detect $k$-qubit errors. However, we consider only the 1-local case throughout the paper, leaving the more general case to the appendix.

%
\section{Error suppression through energy penalties}
%

%
\subsection*{The infinite $\Ep$ case}
%

We first address the question of whether adding an energy penalty works even in principle; that is, we want to show that if $\Ep$ is arbitrarily large, errors are suppressed arbitrarily well. 
Let $U_0(t)$ and $U(t)$ be the evolution operators corresponding to the desired Hamiltonian, $H_0 = \Hsys + \Henv$, and the actual Hamiltonian, $H=H_0 + \lambda V + \Ep \tildeQ$, respectively. That is, $U_0(t)$ and $U(t)$ obey
\begin{eqnarray}
	i \ddt U_0(t) &=& H_0(t) U_0(t), \phantom{\,.}\quad U_0(0)=\id \label{eq:SchroU0}\\
	i \ddt U(t) &=& H(t) U(t), \phantom{_000}\quad U(0)=\id  \notag\,.
\end{eqnarray}
We wish to show that in the codespace, as $\Ep\rightarrow\infty$, $U(t)$ acts as $U_0(t)$. 
Our approach will be to show that the error induced by $V$ is modulated by a term oscillating with frequency $\Ep$ in such a way so that for large $\Ep$ such errors are suppressed.

Our first step is to view $\lambda V$ as a perturbation and to work in the interaction picture using 
\[
	H_{0P}(t) = H_0(t) + \Ep\tildeQ
\]
as the reference Hamiltonian. This corresponds to the evolution operator $U_{0P}(t)$, which obeys
\[
	i \ddt U_{0P}(t) = H_{0P}(t) U_{0P}(t), \quad U_{0P}(0)=\id \,.
\]
Because $H_0$ commutes with each $\P_i$, and therefore with $\tildeQ$, we have that
\begin{equation}\label{eq:U0P}
	U_{0P}(t) = U_0(t)U_P(t) \,,
\end{equation}
where the evolution operator due to the error penalty alone is
\begin{equation*}
	U_P(t) = e^{-i\Ep\tildeQ t} \,.
\end{equation*}
Now, the interaction picture evolution operator
\[
	U_I \equiv U_{0P}^\dag U
\]
obeys
\[
	i \ddt U_I = \lambda V_I U_I \,,
\]
where
\begin{equation}\label{eq:VI}
	\lambda V_I(t) = \lambda U_{0P}^\dag(t) V(t) U_{0P}(t) \,.
\end{equation}
These are just the usual interaction picture equations with a reference Hamiltonian $H_{0P}$ and a perturbation $\lambda V$.
Taking conjugates, we get
\begin{equation}\label{eq:UI}
	U_I^\dag = U^\dag U_{0P} = U^\dag U_0 U_P 
\end{equation}
and
\begin{equation}\label{eq:SchrUI}
	\ddt U_I^\dag = i\lambda U_I^\dag V_I  \,,
\end{equation}
which upon integration gives
\begin{equation} \label{eq:UI_inter}
	U_I^\dag(\tfin) = \id + i\lambda\inttf U_I^\dag V_I \dt \,.
\end{equation}
Note that $\tildeQ\P = 0$, so
\begin{equation}\label{eq:UPP}
	U_P(t) \P = e^{-i\Ep\tildeQ t} \P = \P
\end{equation}
and therefore
\begin{equation*}
	U_I^\dag \P = U^\dag U_0 \P \,.
\end{equation*}
Now, we multiply \eqn{eq:UI_inter} on the right by $\P$, and use this last relation, to get
\begin{equation*}
	U^\dag(\tfin) U_0(\tfin) \P = \P + i\lambda\inttf U_I^\dag V_I \P \dt \,.
\end{equation*}
Multiplying this by $U(\tfin)$ gives
\begin{equation}\label{eq:UIP_inter1}
	U_0(\tfin) \P = U(\tfin) \P + i\lambda U(\tfin) \inttf U_I^\dag V_I \P \dt \,,
\end{equation}
which we can use to track the difference between the evolutions (in the codespace) with and without the coupling to the environment. Our goal is to show that as $\Ep$ goes to infinity, this difference goes to zero. To this end, let
\begin{equation}\label{eq:Fdefn}
	F(t) = \int_0^t V_I(\tau) \P \dd\tau \,.
\end{equation}
Using integration by parts, we see that
\begin{equation*}
\begin{split}
	\inttf U_I^\dag V_I \P \dt
	= \inttf U_I^\dag \ddt[F] \dt
	&= U_I^\dag(\tfin) F(\tfin) - \inttf \ddt[U_I^\dag] F \dt \\
	&= U_I^\dag(\tfin) F(\tfin) - i\lambda \inttf U_I^\dag V_I F \dt \,,
\end{split}
\end{equation*}
where \eqn{eq:SchrUI} was used to obtain the final equality. 
Applying \eqns{eq:UI} and \eqref{eq:VI} we can write this as
\[
	\inttf U_I^\dag V_I \P \dt
		= U^\dag (\tfin) U_{0P}(\tfin) F(\tfin) - i\lambda \inttf U^\dag V U_{0P} F \dt \,
\]
and using this in \eqn{eq:UIP_inter1} we find that
\begin{equation}\label{eq:UIP_main}
\boxedaround{
	U (\tfin) \P
	= U_0(\tfin) \P - i\lambda\Big[ U_{0P}(\tfin) F(\tfin) - i\lambda U(\tfin) \inttf U^\dag V U_{0P} F \dt\Big] \,,
}
\end{equation}
which is an exact expression and not just an expansion in $\lambda$.

We now focus on the operator $F(t)$ defined in \eqn{eq:Fdefn}, which using \eqn{eq:VI} for $V_I$ and \eqn{eq:U0P} for $U_{0P}$ is
\[
	F(t) = \int_0^t U_{P}^\dag U_{0}^\dag V U_{0} U_{P} \P \dd\tau \,.
\]
$\P$ commutes with $H_0$, and therefore also with $U_0$. Because of this and \eqn{eq:UPP} we have
\[
	F(t) = \int_0^t U_0^\dag e^{i\Ep \tildeQ \tau} V \P U_0 \dd\tau \,.
\]
Consider
\begin{equation}\label{eq:eVP_inter1}
	e^{i\Ep \tildeQ \tau} V \P = e^{i\Ep \tildeQ \tau} (V_1 + \cdots + V_{\nlogical}) \P
\end{equation}
where we have written $V$ as the sum over terms associated with each logical qubit, as in \eqn{eq:Vsum}. From the definitions in \eqns{eq:Pprod} and \eqref{eq:tildeQ}, the first term is 
\[
	e^{i\Ep \tildeQ\tau} V_1 \P = e^{i\Ep Q_1 \tau} e^{i\Ep Q_2 \tau} \cdots e^{i\Ep Q_\nlogical \tau} V_1 \P_1 \cdots \P_\nlogical \,.
\]
But, $\P_2\P_3\cdots\P_\nlogical$ commutes with $V_1$, and $\P_i \Q_i = 0$ for all $i$, so we get
\begin{equation}\label{eq:eVP_inter2}
	e^{i\Ep \tildeQ\tau} V_1 \P = e^{i\Ep Q_1 \tau} V_1 \P_1\cdots\P_\nlogical  \,.
\end{equation}
Our code protects against single-qubit errors and we are assuming that the coupling to the environment involves only single-qubit terms, so again,
\[
	\P_1 V_1 \P_1 = 0
\]
which implies that
\begin{equation}\label{eq:VP_QVP}
	V_1 \P_1 = \Q_1 V_1 \P_1 \,.
\end{equation}
Because $\Q_1$ is a projector, we have that
\begin{equation}\label{eq:eVP_inter3}
	e^{i\Ep \Q_1 \tau} \Q_1 = e^{i\Ep\tau} \Q_1 \,.
\end{equation}
The previous equations combine to give
\[
	e^{i\Ep \tildeQ\tau} V_1 \P = e^{i\Ep \tau} V_1 \P
\]
and accordingly,
\begin{equation}\label{eq:UpVP}
	e^{i\Ep \tildeQ \tau} V \P = e^{i\Ep \tau} V \P \,.
\end{equation}
Returning to $F(t)$, we thus have
\begin{equation}\label{eq:Fmodulated}
\boxedaround{
	F(t) = \int_0^t e^{i\Ep \tau} U_0^\dag(\tau) V(\tau) U_0(\tau) \P \dd\tau \,.
}
\end{equation}

Observe that $U_0^\dag(\tau) V(\tau) U_0(\tau) \P$ is independent of $\Ep$. Therefore, we see that when $\Ep$ is large, the integrand of $F$ is a rapidly oscillating function of $\tau$ due to the $e^{i\Ep \tau}$. We can apply the Riemann-Lebesgue lemma to conclude that $F$ vanishes as $\Ep$ goes to infinity. To be explicit, we perform an integration by parts to see that
\begin{eqnarray}\label{eq:Fbyparts}
	F(t) 
	&=& \int_0^t e^{i\Ep \tau} U_0^\dag V U_0 \P \dd\tau \notag\\
	&=& \frac{1}{i\Ep} \Big[ 
		e^{i\Ep t} U_0^\dag (t) V(t) U_0(t) - V(0) - \int_0^t e^{i\Ep \tau} \frac{\dd}{\dd\tau} (U_0^\dag V U_0) \dd\tau
		\Big] \P \,.
\end{eqnarray}
The first two terms in the brackets do not grow with $\Ep$ and the third is bounded by $t$ times the maximum magnitude of $\frac{\dd}{\dd\tau}(U_0^\dag V U_0)$ which is independent of $\Ep$. So as $\Ep$ goes to infinity, $F(t)$ goes to zero. Since both terms in the brackets in \eqn{eq:UIP_main} contain $F$ and are otherwise bounded independent of $\Ep$, we have our $\Ep$ goes to infinity result:
\newtheorem*{onlytheorem}{Theorem} 
\begin{onlytheorem} 
	Suppose that the Hamiltonian of a system coupled to an environment is $$H=\Hsys + \Henv + \lambda V + \Ep\tildeQ,$$ where $V$ acts 1-locally on the system, $\Hsys$ is encoded in a quantum code that can detect single-qubit errors, and $\tildeQ$ is the operator defined in \eqn{eq:tildeQ}. Then, in the limit of an infinitely large energy penalty (positive or negative), the actual evolution in the codespace is as if there were no errors due to $V$; 
i.e. for any time $\tfin$,
\[
\boxedaround{
	\lim_{\Ep\rightarrow\pm\infty} U(\tfin)P = U_0(\tfin) P \,,
}
\]
where $U$ and $U_0$ are the actual and error-free evolution operators defined in \eqn{eq:SchroU0} and $\P$ is the codespace projection operator of \eqn{eq:Pprod}.
\end{onlytheorem}

This result applies to the evolution of both the system and the environment, and is therefore stronger than what we need, which is only that the system evolution be protected. We view our infinite $\Ep$ result as the starting point for large but finite $\Ep$ investigations.

Although throughout this paper we have focused only on the simplest case, where $V$ acts 1-locally on the system and a 1-qubit quantum error detection code is used, this simplification is not necessary. The theorem still holds as long as the error detecting code can detect the errors that $V$ causes, i.e. as long as $PVP=0$, and therefore includes cases where $V$ acts $k$-locally as long as the code can detect $k$-local errors. We show a proof of this in the appendix. The remainder of the paper addresses the case where $V$ acts 1-locally but in which we use a finite, rather than infinite, penalty $\Ep$.

%
\subsection*{The finite $\Ep$ case}
%

\subsubsection*{Frequency Analysis}
We have just seen that for infinitely large $\Ep$, the evolution in the codespace in the presence of noise is the same as the desired noise-free evolution. We now want to know how large $\Ep$ must be to assure us that $F(t)$ is very small, so that the actual evolution in the codespace is close to the desired one. It is helpful to consider the ``natural frequencies" present in the expression for $F(t)$, as given by \eqn{eq:Fmodulated}, which we informally analyze now.

If $f(t)$ is a (suitably nice) complex function, and $\tilde f(\omega)$ is its Fourier transfer, then
\[
	\int_0^t \dd\tau e^{i\Ep\tau} f(\tau)  
	= \int_0^t \dd\tau e^{i\Ep\tau} \int_{-\infty}^\infty \dd\omega e^{-i\omega\tau} \tilde f(\omega)
	= \int_{-\infty}^\infty \dd\omega \frac{e^{i(\Ep-\omega)t} -1}{i(\Ep-\omega)} \tilde f(\omega) \,.
\]
Suppose that there exists an $\omega_c$ such that $\tilde f(\omega)$ is non-negligible only for $|\omega| < \omega_c$.
Then  
$$
	\int_0^t \dd\tau e^{i\Ep\tau} f(\tau)
	\approx
	\int_{|\omega| \leq \omega_c} \dd\omega \frac{e^{i(\Ep-\omega)t} -1}{i(\Ep-\omega)} \tilde f(\omega) \,.
$$ 
Now, if $\Ep$ is much larger than $\omega_c$, we can replace $1/(\Ep-\omega)$ by $1/\Ep$ in this integral, and may therefore conclude that the integral is small (shrinking as $1/\Ep)$.

The question is, therefore, what are the natural frequencies of $U_0^\dag(\tau) V(\tau) U_0(\tau)$? If they are not too large, then $F(t)$ should be small for reasonably large values of $\Ep$.
Consider first the time-independent case, in which $H_0$ and $V$ are time-independent, so $U_0(\tau) = e^{-iH_0 \tau}$. Certainly $U_0$ will have extremely large frequencies, namely $e^{-iE\tau}$ where $E$ are eigenenergies of $H_0$; since $H_0$ includes the environment, $E$ can scale with the size of the environment and be extremely large. However, the frequencies of $U_0^\dag V U_0$, are expected to be much smaller. Inserting two complete sets of $H_0$ energy eigenstates, $\ket{E}$, we see that in the time-independent case,
$$
	U_0^\dag(\tau) V U_0(\tau) = \sum_{E,E'} e^{i(E - E')\tau} \ket{E} \bra{E} V \ket{E'} \bra{E'} \,,
$$
indicating that the frequencies are the energy \textit{differences} induced by $V$.
If $V$ acts locally, we expect it would be unable to change the energy of the system/environment by a large amount - for example, it is unlikely that flipping just two spins in a spin chain will change the energy of the entire chain by more than a small amount. Therefore, we expect that $\bra{E} V \ket{E'}$ is very small when $|E-E'|$ is large. If we make $\Ep$ larger than the largest $|E-E'|$ corresponding to any non-negligible $\bra{E} V \ket{E'}$, we can conclude that $F$ is small. To be more precise would require a specific model for the system, environment, and interaction. Still, we can make some progress on bounding $F$, even in the general time-dependent case.

\subsubsection*{Bounding $F$}
We now bound the norm of $F(t) = \int_0^t e^{i\Ep \tau} U_0^\dag(\tau) V(\tau) U_0(\tau) \P \dd\tau$. 
Since the norm of $V$ is expected to grow linearly in the size of the system, and therefore in $\nlogical$, one would naively expect the same of $F$. However, the fact that each logical qubit is independently encoded allows us to do slightly better.
Recall from \eqn{eq:Vsum} that we can write $V$ as a sum of terms, $V_i$, where each $V_i$ acts only on the $i$th logical system qubit (as well as the environment). Let
$$ 
	F_i(t) = \int_0^t e^{i\Ep \tau} U_0^\dag(\tau) V_i(\tau) U_0(\tau) \P \dd\tau 
$$
so 
$$	
	F = \sum_{i=1}^\nlogical F_i \,.
$$
We now show that
\begin{equation}\label{eq:FisFisum}
	\norm{F} \leq \sqrt{\nlogical} \max_i \norm{F_i} \,.
\end{equation}

\begin{proof}
Observe that 
\begin{eqnarray*}
F^\dag F
&=&
\sum_{i,j=1}^\nlogical F_i^\dag F_j \\
&=&
 \sum_{i,j=1}^\nlogical \int_0^t \int_0^t \dd\tau_1 \dd\tau_2 e^{i\Ep (\tau_2-\tau_1)} [\P U_0^\dag(\tau_1) V_i(\tau_1) U_0(\tau_1)]   [U_0^\dag(\tau_2) V_j(\tau_2) U_0(\tau_2) \P] \,
\end{eqnarray*}
and consider the terms with $i\neq j$. In the first $\P$ there is a $\P_j$ (i.e. $\P = \P \P_j$) and it commutes with $U_0^\dag(\tau_1), V_i(\tau_1)$, $U_0(\tau_1)$, and $U_0^\dag(\tau_2)$. But $\P_j V_j \P = 0$ so these terms are 0. Consequently, the sum is only over $i=j$, i.e.
\[
	F^\dag F = \sum_{i=1}^\nlogical F_i^\dag F_i \,,
\]
and the claim follows since $\norm{F}^2 = \max_{\ketpsi} \brapsi F^\dag F \ketpsi$.
\end{proof}

We now consider how to bound $F_i$ (for any logical qubit $i$).
In deriving \eqn{eq:Fbyparts}, we assumed that $\frac{\dd}{\dd\tau}(U_0^\dag V U_0)$ is finite; we now explicitly bound this term.
By \eqn{eq:SchroU0}, we have 
$$
	\frac{\dd}{\dd\tau}(U_0^\dag V_i U_0) = -i U_0^\dag [V_i,H_0] U_0 + U_0^\dag\frac{\dd V_i}{\dd\tau}U_0 \,.
$$
Using this, \eqn{eq:Fbyparts} becomes
\begin{equation*}
	F_i(t) 
	= \frac{1}{i\Ep} \Big[ 
			e^{i\Ep t} U_0^\dag (t) V_i(t) U_0(t) 
			- V_i(0) 
			+i  \int_0^t e^{i\Ep \tau} U_0^\dag [V_i,H_0] U_0 \dd\tau
			- \int_0^t e^{i\Ep \tau} U_0^\dag\frac{\dd V_i}{\dd\tau}U_0 \dd\tau
		\Big] \P  
\end{equation*}
and taking the norm, using that $\norm{A+B}\leq\norm{A}+\norm{B}$, $\norm{AB}\leq\norm{A}\norm{B}$, $\norm{U_0}=1$, and $\norm{P}\leq 1$, we obtain
\begin{equation}\label{eq:Fcomm_norm_with_deriv}
\boxedaround{
	\bignorm{F_i(t)} \leq 
	\frac{1}{|\Ep|}\left(\Big\Vert{V_i(t)}\Big\Vert + \Big\Vert{V_i(0)}\Big\Vert + \max_\tau \Big\Vert{[V_i(\tau),H_0(\tau)]}\Big\Vert t 
	+ \max_\tau \norm{\frac{\dd V_i}{\dd\tau}}t 
	\right) \,.
}
\end{equation}

The norm $\norm{\frac{\dd V_i}{\dd\tau}}$
will be bounded for reasonable $V$. For example, if the system control operations do not greatly change the environment surrounding each qubit, one expects that each $V_i$ will likely stay fairly constant. Accordingly, we will ignore this term and the time dependence of $V_i$,
in which case
\begin{equation}\label{eq:Fcomm_norm}
	\bignorm{F_i(t)} \leq 
	\frac{1}{|\Ep|}\left(2\Big\Vert{V_i}\Big\Vert + \max_\tau \Big\Vert{[V_i,H_0(\tau)]}\Big\Vert t	\right) \,.
\end{equation}
The commutator $[V_i,H_0]=[V_i,\Hsys + \Henv]$ involves the environment Hamiltonian, which may be extremely large; however, we now show that by making some reasonable physical assumptions, $\bignorm{[V_i,H_0]}$ is independent of the size of the system and environment.

First, we assume that $\Hsys, \Henv$, and $V_i$ are local operators. They can therefore each be written as a sum of terms, each term involving only a few qubits.  Second, we make the assumption that each qubit (of the system and environment) appears in at most a few of these local terms of $\Hsys, \Henv$, and $V_i$. 
For example, if a Hamiltonian is 2-local and \textit{geometrically} local, say on a cubic lattice, so that each qubit only interacts with its immediate neighbors, then this restricts the number of terms in which any qubit appears, say to six for the cubic lattice. 
In terms of operator norms, these assumptions translate as follows.

For $V_i$ we have
\[
	V_i 
	= \sum_{\physdummy=1}^{\nphysperlog} \sum_{\paulidummy=X,Y,Z} \sigma_\paulidummy^\physdummy \otimes B_\paulidummy^\physdummy
\]
where the sum over $\physdummy$ is only over the $\nphysperlog$ system qubits that comprise the $i$th logical qubit. $B_\paulidummy^\physdummy$ is an environmental operator that couples to $\sigma_\paulidummy^\physdummy$ and only consists of a few local terms (because system qubit $\physdummy$ only appears in a few local terms of $V_i$), each acting on only a few environmental qubits (by locality). Therefore, 
\[
	\norm{B_\paulidummy^\physdummy} = \bigO{1}
\]
independent of the system and environment sizes. (Recall that the coupling, $\lambda$, has units of energy, so the $B_\paulidummy^\physdummy$ are dimensionless.) We thus have that 
\begin{equation}\label{eq:Vi_bound}
	\norm{V_i} = \nphysperlog \bigO{1} \,.
\end{equation}

\newcommand{\hsysa}{h_\text{sys}^\physdummy}
\newcommand{\henva}{h_\text{env}^\physdummy}
Now, $H_0 = \Hsys + \Henv$ and both terms contribute to the commutator $[V_i,H_0]$. 
Let $\hsysa$ be the sum of all terms in $\Hsys$ involving system qubit $\physdummy$, where $\physdummy$ is a part of logical qubit $i$. Since there are only a few such terms, each of which acts on only a few system qubits, we can assert that
\[
	\norm{\hsysa} =  \calE \bigO{1}
\]
where $\calE$ is an energy scale parameter whose size is on the order of the size of the individual terms in $\Hsys$. Similarly, let $\henva$ be the sum of all terms in $\Henv$ that contain the environmental qubits that appear in $B_\paulidummy^\physdummy$ for $\paulidummy=X,Y,Z$. Since $B_\paulidummy^\physdummy$ involves only a few environment qubits, which each appear in $\Henv$ in only a few, local terms, we have that
\[
	\norm{\henva} =  \calE \bigO{1} \,.
\]
Then, since $\norm{A+B}\leq\norm{A}+\norm{B}$, $\norm{[A,B]}\leq 2\norm{A}\norm{B}$, and $\norm{A\otimes B}=\norm{A}\norm{B}$,
\[
\begin{split}
	\Big\Vert{[V_i,H_0]}\Big\Vert 
	&\leq
	\sum_{\physdummy=1}^{\nphysperlog} \sum_{\paulidummy=X,Y,Z} 
		\Big\Vert{[\sigma_\paulidummy^\physdummy \otimes B_\paulidummy^\physdummy,\Hsys]}\Big\Vert 
		+ \Big\Vert{[\sigma_\paulidummy^\physdummy \otimes B_\paulidummy^\physdummy,\Henv]}\Big\Vert \\
	&=
	\sum_{\physdummy=1}^{\nphysperlog} \sum_{\paulidummy=X,Y,Z} 
		\Big\Vert{[\sigma_\paulidummy^\physdummy \otimes B_\paulidummy^\physdummy,\hsysa]}\Big\Vert + 
		\Big\Vert{[\sigma_\paulidummy^\physdummy \otimes B_\paulidummy^\physdummy,\henva]}\Big\Vert \\
	&\leq
	2\sum_{\physdummy=1}^{\nphysperlog} \sum_{\paulidummy=X,Y,Z}
		 \norm{\sigma_\paulidummy^\physdummy}\norm{B_\paulidummy^\physdummy} 
		 \left(\norm{\hsysa} + \norm{\henva} \right) \,.
\end{split}
\]
Thus,
\begin{equation}\label{eq:VH_bound}
	\Big\Vert{[V_i,H_0]}\Big\Vert =  \nphysperlog\calE \bigO{1} \,
\end{equation}
independent of $\nlogical$ and the size of the environment.

Applying the bounds of \eqns{eq:Vi_bound} and \eqref{eq:VH_bound} to \eqn{eq:Fcomm_norm} gives
\[
	\norm{F_i(t)} \leq  \frac{1}{|\Ep|} \big[ \nphysperlog \bigO{1} +  \nphysperlog \calE t\bigO{1} \big] \,
\]
and using this in \eqn{eq:FisFisum}, we obtain
\begin{equation}\label{eq:F_bound}
	\norm{F(t)} \leq  \frac{\sqrt{\nlogical}}{|\Ep|} \nphysperlog \big[\bigO{1} +   \calE t\bigO{1} \big] \,.
\end{equation}
The term that grows with $t$ represents a very weak bound for large $t$. We see from \eqn{eq:Fmodulated} that $F(t)$ is an integral over $[0,t]$ of an oscillating integrand and such integrals typically do not grow with $t$. For example, bounding
\[
	\int_0^t \sin(\omega \tau) \dtau \leq t \,,
\]
while true, is not very helpful for large $t$. However, this is the best that we have been able to do for the general problem at hand. In Sec.~\ref{sec:numerics} we will look at the full $t$ dependence of small systems using numerical simulation.

\subsubsection*{Fidelity calculation}
\newcommand{\fid}{\calF}
Suppose the system/environment is initially in the pure state $\ketpsi$, and it evolves under $U$ for time $\tfin$. We begin in the codespace of the system, so $\P\ketpsi = \ketpsi$.
The fidelity squared, $\fid^2$, between the desired final state, $U_0\ketpsi$, and the actual final state, $U\ketpsi$, is given by
\[
	\fid^2 
	= \left| \bra{\psi} U_0^\dag U \ket{\psi} \right|^2 
	= \left| \bra{\psi} \P U_0^\dag U \P \ket{\psi} \right|^2 \,.
\]
To evaluate this, we left-multiply \eqn{eq:UIP_main} by $\P U_0^\dag$, and use \eqn{eq:UPP} to give
\[
	\P U_0^\dag U \P
	= \P  - i\lambda \P F - \lambda^2 \P U_0^\dag U \int_0^{\tfin} U^\dag V U_{0P} F \dt \,.
\]
Because $\P$ commutes with $U_0$ and $PVP=0$, we see from \eqn{eq:Fmodulated} that $\P F = 0$. Therefore,
\[
	\P U_0^\dag U \P = \P - \lambda^2 \P U_0^\dag U \inttf U^\dag V U_{0P} F \dt \,,
\]
so
\begin{equation}\label{eq:overlap}
	\bra{\psi} U_0^\dag U \ket{\psi} 
	= 1 - \lambda^2 \brapsi \P U_0^\dag U \inttf U^\dag V U_{0P} F \dd\tau \ketpsi\,.
\end{equation}

We assume that $\lambda$, the system-environment coupling, can be engineered to be small compared to the magnitudes of the individual terms in $H_0$. Accordingly, let us consider $\bra{\psi} U_0^\dag U \ket{\psi}$ to order $\lambda^2$. Working to this order, we can set $U = U_{0P}$ (as would occur if $\lambda$ were zero) on the right hand side of \eqn{eq:overlap}:
\[
	\bra{\psi} U_0^\dag U \ket{\psi} 
	= 1 - \lambda^2 \brapsi \P U_0^\dag U_{0P} \inttf U_{0P}^\dag V U_{0P} F \dt \ketpsi\,  		+ \,\bigO{\lambda^3}.
\]
Recall from \eqns{eq:U0P} and \eqref{eq:UPP} that $U_{0P} = U_0 U_P$ and $P U_P = P$, so that $\P U_0^\dag U_{0P} = \P$. 
Recalling the notation of \eqn{eq:VI} from the interaction picture, i.e. of $V_I \equiv U_{0P}^\dag V U_{0P}$, and the definition of $F$ in \eqn{eq:Fdefn}, we therefore have
\[
\begin{split}
	\bra{\psi} U_0^\dag U \ket{\psi}
	&= 1 - \lambda^2 \brapsi \P \inttf V_I(t) F(t) \dt \ketpsi\,  		+ \,\bigO{\lambda^3} \\
	&= 1 - \lambda^2 \brapsi \P \inttf V_I(t) \int_0^t V_I(\tau) \P \ \dtau \dt \ketpsi\,  		+ \,\bigO{\lambda^3}.
\end{split}
\]

With perfect error suppression, $\fid^2 \rightarrow 1$, so $1-\fid^2$ is a measure of error suppression failure. We calculate
\[
	1 - \fid^2 
	= 1 - \left| \bra{\psi} U_0^\dag U \ket{\psi} \right|^2 
	= \lambda^2 \brapsi\P \inttf  \dt \int_0^t \dtau \  V_I(t)V_I(\tau) \P\ketpsi + { h.c. } + \bigO{\lambda^3}
\]
where \textit{h.c.} denotes the Hermitian conjugate. But this conjugate involves
\[
	\left( \inttf  \dt \int_0^t \dtau \  V_I(t)V_I(\tau) \right)^\dag
	= \inttf  \dt \int_0^t \dtau \  V_I^\dag(\tau) V_I^\dag(t)
	= \inttf  \dtau \int_0^\tau \dt \  V_I(t) V_I(\tau)
\]
where in the last step we used the fact that $V_I$ is Hermitian and relabeled $t \leftrightarrow \tau$, showing that this term is identical to the to the term of which it is the conjugate, except for the integration region. The original integrates over a region with $\tau<t$, while the conjugate integrates the same integrand over a region with $t<\tau$, so their sum integrates over all $0\leq \tau, t \leq \tfin$. 
Thus,
\[
	1 - \fid^2 
	= \lambda^2 \brapsi\P \inttf \dt V_I(t) \inttf \dtau V_I(\tau)  \P\ketpsi + \, \bigO{\lambda^3}
\]
i.e.
\[
	1 - \fid^2 
	= \lambda^2 \brapsi F^\dag F \ketpsi + \,\bigO{\lambda^3}
\]
so
\begin{equation}\label{eq:fid_bound_lambda2}
\boxedaround{
	1-\fid^2 \leq \lambda^2 \norm{F}^2 + \,\bigO{\lambda^3} \,.
}
\end{equation}
We see that a small $\norm{F}$ corresponds to good error suppression.

We can combine this expression with \eqn{eq:F_bound} to obtain, at time $\tfin$,
\[
	1-\fid^2(\tfin) \leq \frac{\lambda^2 {\nlogical}}{\Ep^2} \nphysperlog^2 \big[\bigO{1} +   \bigO{1} \calE \tfin  \big]^2
		+ \, \bigO{\lambda^3} \,.
\]
It is possible to write an expression for the $\lambda^3$ contribution. We find that the leading term in $1/\Ep$ in the $\lambda^3$ contribution goes like $\lambda^3\tfin/\Ep^2$. Again, we do not believe that this gives a useful bound for large $\tfin$, but it may be useful in the small $\tfin$ regime.

%
\section{Numerical simulation for one logical qubit} \label{sec:numerics}
%

\newcommand{\nenv}{n_e}
\newcommand{\nenvactual}{8}
\newcommand{\nphysicalactual}{4}
In this section, we discuss the results of a numerical simulation of
one logical qubit, encoded as 4 physical qubits using the Jordan-Farhi-Shor\cite{JFS06} code, coupled to an $\nenvactual$-qubit environment according to
\[
	H = \Hsys + \Henv + \lambda V + \Ep \tildeQ \,.
\]
Since we track the evolution over long times, we find it too computationally expensive to work with more than 12 qubits total; therefore, for this paper we analyze only one logical qubit coupled to a modest size environment.

We choose the environment and the couplings as follows.
The environment qubits are arranged on a randomly chosen 3-regular graph and have 2-local interactions between nearest neighbors. Each physical system qubit couples to a single, unique, randomly-selected environment qubit. For simplicity, the environment and coupling Hamiltonians, $\Henv$ and $V$, are time-independent.

\newcommand{\envduma}{a}
\newcommand{\envdumb}{b}
\newcommand{\envdumc}{c}
\newcommand{\ndotsigma}[2]{(\hat #1_{#2} \cdot \vec \sigma^{#2})}
We choose the environment Hamiltonian to be
\[
	\Henv = 	
		\sum_{\envduma=1}^{\nenvactual} \alpha_\envduma \ndotsigma{n}{\envduma}
		+  \sum_{\langle \envdumb,\envdumc \rangle} \alpha_{\envdumb\envdumc} \ndotsigma{m}{\envdumb} \otimes \ndotsigma{\ell}{\envdumc} 
\]
where each $\hat n_\envduma, \hat m_\envdumb$, and $\hat \ell_\envdumc$ is a randomly chosen unit vector, $\vec \sigma^\envduma = (\sigma_X^\envduma,\sigma_Y^\envduma,\sigma_Z^\envduma)$ are the Pauli operators acting on environment qubit $\envduma$, each $\alpha_\envduma$ and $\alpha_{\envdumb\envdumc}$ is a coefficient chosen at random in the range of $[0.9,1.1]$, and $\sum_{\langle \envdumb,\envdumc \rangle}$ denotes a sum over neighboring environment qubits on the 3-regular graph.

In this small simulation, with one logical qubit, the system size is 4.
The system-environment coupling has the form of \eqn{eq:Vdefn}, with the environmental operators chosen to be simple single-qubit terms, and is given by
\[
	V = 
		\sum_{\physdummy=1}^{\nphysicalactual} \beta_\physdummy  \ndotsigma{n}{\physdummy} 
		+  \sum_{\physdummy=1}^\nphysicalactual \gamma_{\physdummy} \ndotsigma{m}{\physdummy} \otimes  (\hat \ell_{\physdummy} \cdot \vec \sigma^{\physdummy}_\text{env})
\]
where each $\hat n_{\physdummy}$, $\hat m_{\physdummy}$, and $\hat \ell_{\physdummy}$ is a randomly chosen unit vector, $\vec \sigma^{\physdummy}$ are the Pauli operators acting on system qubit $\physdummy$, and $\vec \sigma^{\physdummy}_\text{env}$ are the Pauli operators acting on the environment qubit that is coupled to system qubit ${\physdummy}$. Note that we have included single-qubit error terms, $\hat n_{\physdummy} \cdot \vec \sigma^{\physdummy}$, that are not coupled to any environment qubits but may arise from pure system errors.
The coefficients $\beta_\physdummy$ and $\gamma_{\physdummy}$ are each chosen at random in the range of $[0.9,1.1]$.
By design, $V$ acts 1-locally on the system.

The initial state is taken to be a pure product state of the system and environment,
\[
	\ketpsi = \ket{\psi^s} \otimes \ket{\psi^e} \,,
\]
where the initial environment state $\ket{\psi^e}$ is a random $\nenvactual$-qubit state. We will study different choices for the initial system state $\ket{\psi^s}$ and the computational Hamiltonian $\Hsys$. In order to compare the actual and desired dynamics, we evolve with $U$ and $U_0$ defined in \eqn{eq:SchroU0} to obtain
\[
\begin{split}
	\ket{\phi(t)} &= U(t)\ketpsi,  \phantom{_0}\qquad t\in [0,\tfin] \\
	\ket{\phi_0(t)} &= U_0(t)\ketpsi, \qquad t\in [0,\tfin] \,.
\end{split}
\]
Note that because the system and environment are not coupled by $H_0$, we can write $$\ket{\phi_0(t)} = \ket{\phi_0^s(t)} \otimes \ket{\phi_0^e(t)}$$ so that the state of the system at time $t$ is $\ket{\phi_0^s(t)}$, independent of the environment. In the coupled case, on the other hand, the state of the system at time $t>0$ is described by a density matrix,
$$\rho(t) = \tr_\text{env} \selfketbra{\phi(t)},$$
where the environment qubits have been traced out.

At any time $t$, we compare the actual versus coupling-free evolutions using the following measures:
\begin{itemize}
	\item
		The squared fidelity of the total evolution, $$\fid^2(t) = |\braket{\phi_0(t)}{\phi(t)}|^2.$$
		As a result of our theorem, this measure goes to 1 as $\Ep\rightarrow\pm\infty$.  This fidelity also contains the fidelity of the environment's evolution, and accordingly is a stronger measure than what we need to track how well the computation is protected.
		
	\item
		The squared fidelity of the \textit{system} evolution, $$\fid_s^2(t) = \bra{\phi_0^s(t)} \rho(t) \ket{\phi_0^s(t)}.$$ 
		This measure determines if the quantum computation in the presence of the coupling to the environment is following the desired evolution. The irrelevant bath degrees of freedom are traced out.
\end{itemize}

\begin{figure} 
\begin{center}
\placefidplot{{fs_t_rand}}
\placefidplot{{f_t_rand}}
\caption{\label{fig:XL_rand} 
(Top) squared system fidelity, $\fid_s^2$, and (bottom) squared total fidelity, $\fid^2$, 
as functions of time $t$ on a log-scale
for $\lambda=0.1$ and initial system state 
$\ket{\psi^s} = \alpha\ket{+_L} + \beta\ket{-_L}$ with a random choice of $\alpha$ and $\beta$ obeying $|\alpha|^2+|\beta|^2=1$. 
Results are shown for increasing energy penalty strengths, $\Ep$. All data are for $\Hsys = X_L$ on a 4-qubit system, and for a particular random instance of $\Henv, V$, $\ket{\psi^s}$, and $\ket{\psi^e}$ with $\nenvactual$ environment qubits. The dashed line in the top panel is at $|\alpha|^4+|\beta|^4$, which is $0.615$ for this particular choice of $\ket{\psi^s}$; its significance will be explained later. The dashed line at 1/16 is the expected long time system fidelity in the absence of protection. }
\end{center}
\end{figure}

We first perform simulations for the time-independent computational Hamiltonian $\Hsys=X_L$. Figure~\ref{fig:XL_rand} shows the results of a typical simulation with $\lambda=0.1$ and for a variety of $\Ep$ values, for both fidelity measures defined above.  
The initial system state in this case is a random superposition of $\ket{0_L}$ and $\ket{1_L}$, which can be viewed as a random superposition of the codespace eigenstates of $\Hsys = X_L$, i.e. of
$$
	\ket{\pm_L} = \frac{1}{\sqrt2}\Big(\ket{0_L} \pm \ket{1_L}\Big) \,.
$$ 
We make the following observations:

\begin{itemize}
\item
	In the absence of an energy penalty, i.e. when $\Ep=0$, the fidelities rapidly fall. We see that $\fid_s^2$ falls to a value of about $1/16$, which is the expected fidelity between two random 4-qubit system states. In other words, the state of the system is outside the codespace and is uncorrelated with the state resulting from the desired evolution. 
	
\item
	For large $\Ep$, near-perfect fidelity is maintained for a long time, both for the system ($\fid_s$) and the system-environment ($\fid$).	
	However, the fidelity eventually falls, and does so fairly abruptly (on a log-scale). This kind of behavior would not be seen in a low-order power series expansion in time and is certainly not seen in expressions like \eqn{eq:Fcomm_norm_with_deriv} that have a linear term in $t$.
	Note that the larger the value of $\Ep$, the longer near-perfect fidelity is maintained. 

\item
	For sufficiently large $\Ep$, the general behavior is for the system fidelity $\fid_s$ to approach an asymptotic value for large $t$, about which it has small fluctuations.
	We have data for times greater than what we plot here that supports this observation, but of course we cannot draw firm conclusions about what happens as $t \rightarrow\infty$. Still, we can say that the system fidelity stays fairly level away from zero at time scales much larger than any natural time scale involved in the simulation.
	
\item	
	The total fidelity, $\fid$, always falls to very close to 0 for very large $t$, indicating that the environment state is not as well protected as the system state is. This is unsurprising, as there is no preferred codespace for the environment.
	
\item
	In Fig.~\ref{fig:XL_lam0.01}, we see qualitatively the same behavior for the same randomly chosen $\Henv, V$, and $\ket{\psi}$, but with $\lambda=0.01$ (rather than $\lambda=0.1$). Note that for each $\Ep$, the smaller $\lambda$ value allows for good protection for longer times than the larger $\lambda$ value allows. 

	\begin{figure}
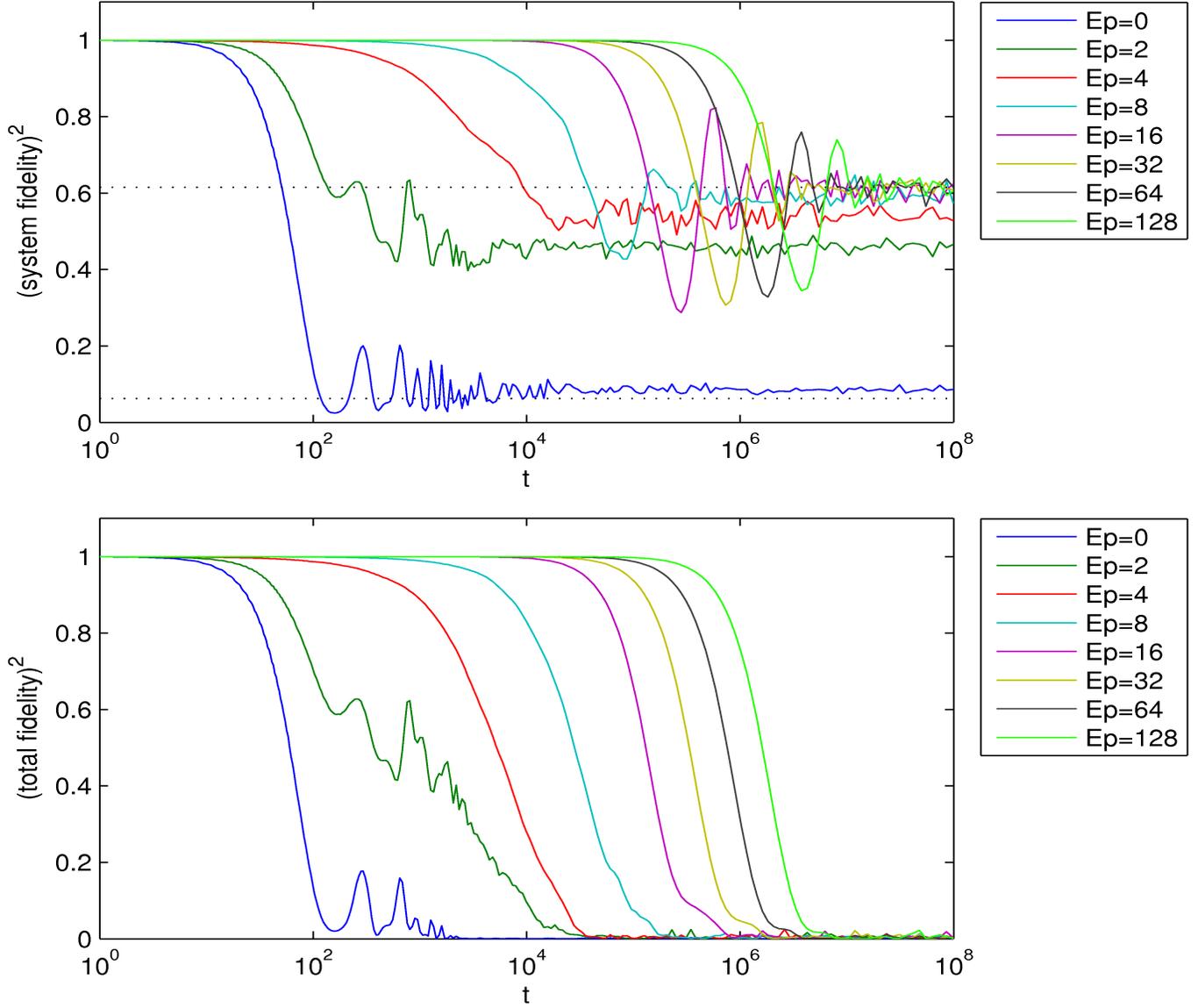
 
	\begin{center}
	\placefidplot{{fs_t_lam001}}
	\placefidplot{{f_t_lam001}}
	\caption{\label{fig:XL_lam0.01} 
	(Top) squared system fidelity, $\fid_s^2$, and (bottom) squared total fidelity, $\fid^2$, 
	as functions of time $t$ on a log-scale
	for $\lambda=0.01$. All other values are identical to those of Fig.~\ref{fig:XL_lam0.01}, but the time scale has been increased because there is better protection for the smaller value of $\lambda$.}
	\end{center}
	\end{figure}
\end{itemize}
	
%
%
It is interesting to compare the bounds of \eqn{eq:fid_bound_lambda2} and \eqn{eq:Fcomm_norm} with our numerical observations. For the parameters used to generate Figs.~\ref{fig:XL_rand} and \ref{fig:XL_lam0.01}, we have $\Vert{V}\Vert \approx 7$, $\Vert{H_0}\Vert \approx 12$, and $\big\Vert{[V,H_0]}\big\Vert \approx 17$ (significantly less than $\Vert{V}\Vert \cdot \Vert{H_0}\Vert$, in accordance with our previous discussion on locality). Equation~\eqref{eq:fid_bound_lambda2} suggests that good fidelity squared, say of 0.9, can be achieved if $\lambda^2 \norm{F}^2 \lesssim 0.1$, so for $\lambda=0.1$ we expect that we need $\norm{F} \lesssim 3$. The bound in \eqn{eq:Fcomm_norm} indicates that for $\Ep=32$, $\norm{F} \lesssim 3$ for $\tfin \lesssim 5$, so that these two bounds together suggest that good fidelity can be maintained for time $\tfin \lesssim 5$ if $\Ep=32$. 
However, in Fig.~\ref{fig:XL_rand} we see that, in this case, we can maintain good $\fid^2$ up to $\tfin=1000$.
Similarly, for $\lambda=0.01$ we expect that we need $\norm{F} \lesssim 30$ (from \eqn{eq:fid_bound_lambda2}), which for $\Ep=32$ can be guaranteed for $\tfin \lesssim 60$ (by \eqn{eq:Fcomm_norm}); however, Fig.~\ref{fig:XL_lam0.01} indicates that in this case we can maintain good $\fid^2$ up to $\tfin=100,000$.
We thus see that \eqn{eq:Fcomm_norm} is not really useful for large $\tfin$, as our numerical results show good fidelity for far longer than our bounds can guarantee.

%
%
\newcommand{\tprot}{t_\text{prot}}
\begin{figure}
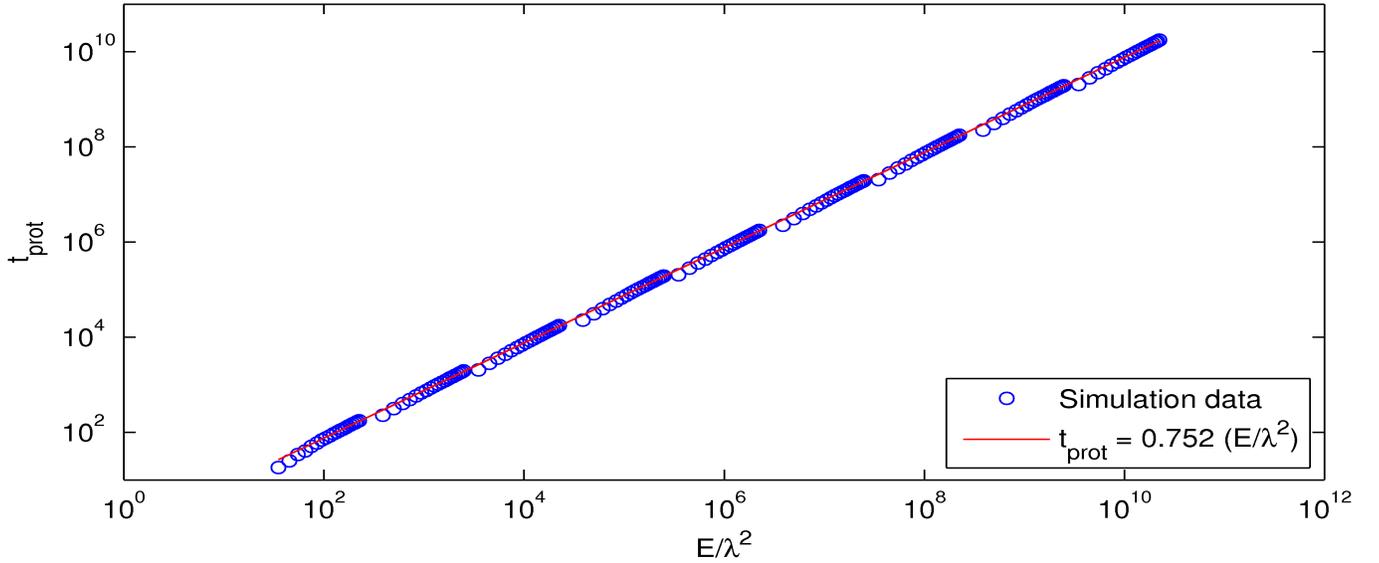
 
\begin{center}
\placefidplot{{protection_time}}
\caption{\label{fig:protection_time} 
The protection time, $\tprot$, defined as the time at which the squared system fidelity $\fid_s^2$ drops to 0.9, versus $\Ep/\lambda^2$ for a range of $\Ep$ and $\lambda$ values, specifically, $\Ep \in \lbrace 35,45,\ldots,225 \rbrace$ and $\lambda \in \lbrace 10^{-4}, 3\cdot 10^{-4}, 10^{-3}, 3\cdot 10^{-3}, 10^{-2}, 3\cdot 10^{-2}, 10^{-1}, 3\cdot 10^{-1}, 1 \rbrace$ (each of the 9 ``clusters" of data in the figure corresponding to a different $\lambda$ value.)
All Hamiltonian and initial state values (other than $\Ep$ and $\lambda$) are kept identical to those of Fig.~\ref{fig:XL_rand}.
The line shows that a linear relationship between $\tprot$ and $\Ep/\lambda^2$ fits the data well.}
\end{center}
\end{figure}

To address the question of how long good fidelity can be maintained, we note that for successful quantum computation it suffices to have high system fidelity $\fid_s$; high total fidelity $\fid$ is not required.
Accordingly, we define the \textit{protection time}, $\tprot$, to be the time at which the squared system fidelity $\fid_s^2$ first drops to 0.9. In Fig.~\ref{fig:protection_time} we plot $\tprot$ for a variety of values of $\Ep \in [35,225]$ and $\lambda \in [10^{-4},1]$ (with all other Hamiltonian and initial state values held fixed). Observe that, to a very good approximation, the data fits the relation 
$$\tprot \propto \Ep/\lambda^2 \,$$
especially for larger values of $\Ep$. We will later show a simple model that is consistent with this behavior.

%
%
\begin{figure}
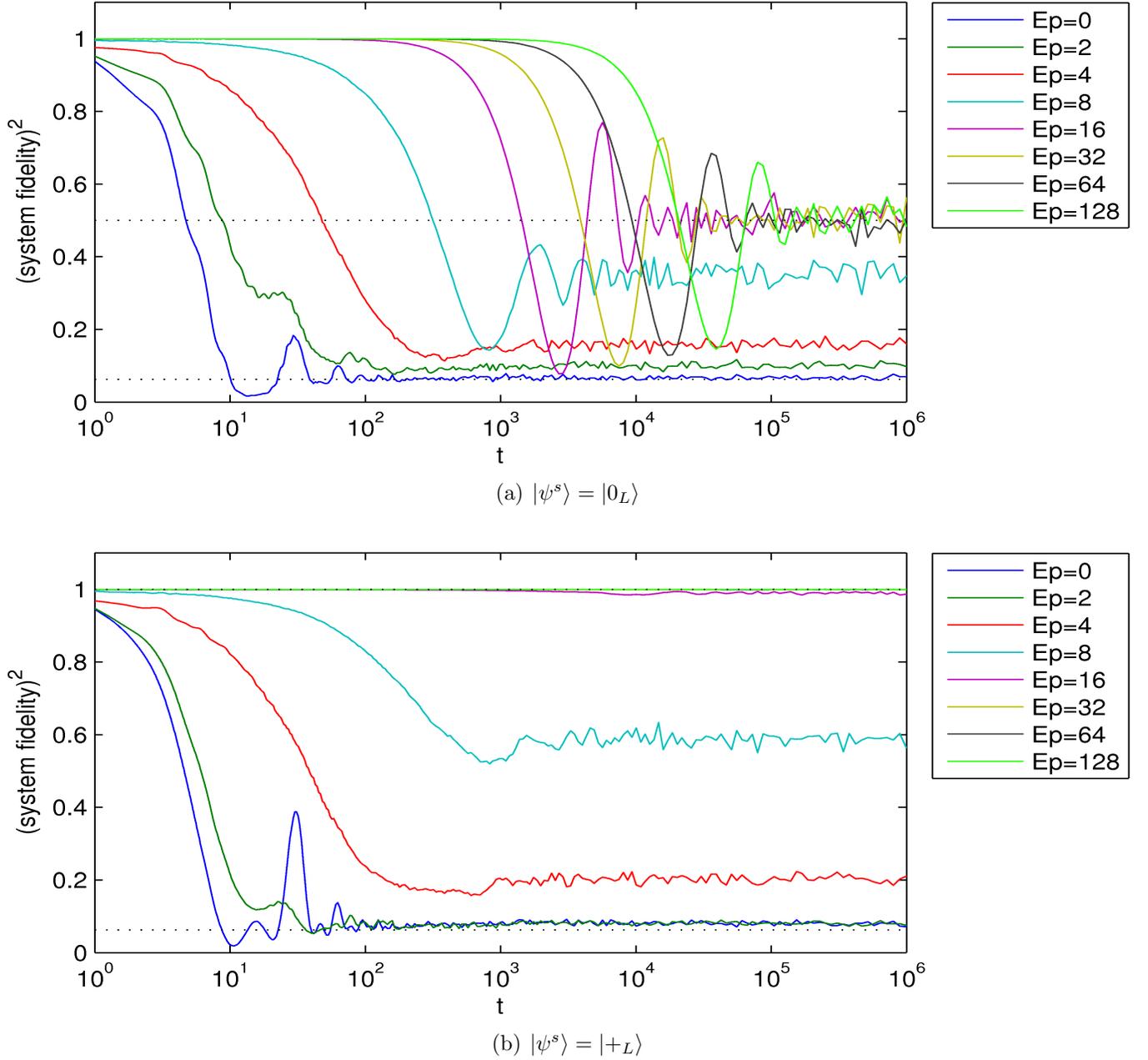
 
\begin{center}
	\subfigure[$\ket{\psi^s}=\ket{0_L}$]
	 {
        \placefidplot{{fs_t_0}}
        \label{fig:XL_0L}
    }
    \\
    \subfigure[$\ket{\psi^s}=\ket{+_L}$]
    {
        \placefidplot{{fs_t_eig}}
        \label{fig:XL_eig}
    }
\caption{ 
Squared system fidelity $\fid_s^2$ as a function of time $t$ on a log-scale 
for initial states
(a) $\ket{\psi^s}=\ket{0_L}$ and
(b) $\ket{\psi^s}=\ket{+_L}$,
with $\Hsys=X_L$ and $\lambda=0.1$. 
All Hamiltonian and initial environment state values are identical to those of Fig.~\ref{fig:XL_rand}. The dashed lines at 1/2 (in the top figure) and 1/16 (in both figures) serve as guides for the eye. Note that in the bottom figure, for $\Ep \geq 16$, $\fid_s^2$ remains close to 1 for the duration of the simulation.}
\label{fig:XL_0Landeig}
\end{center}
\end{figure}

\begin{figure}
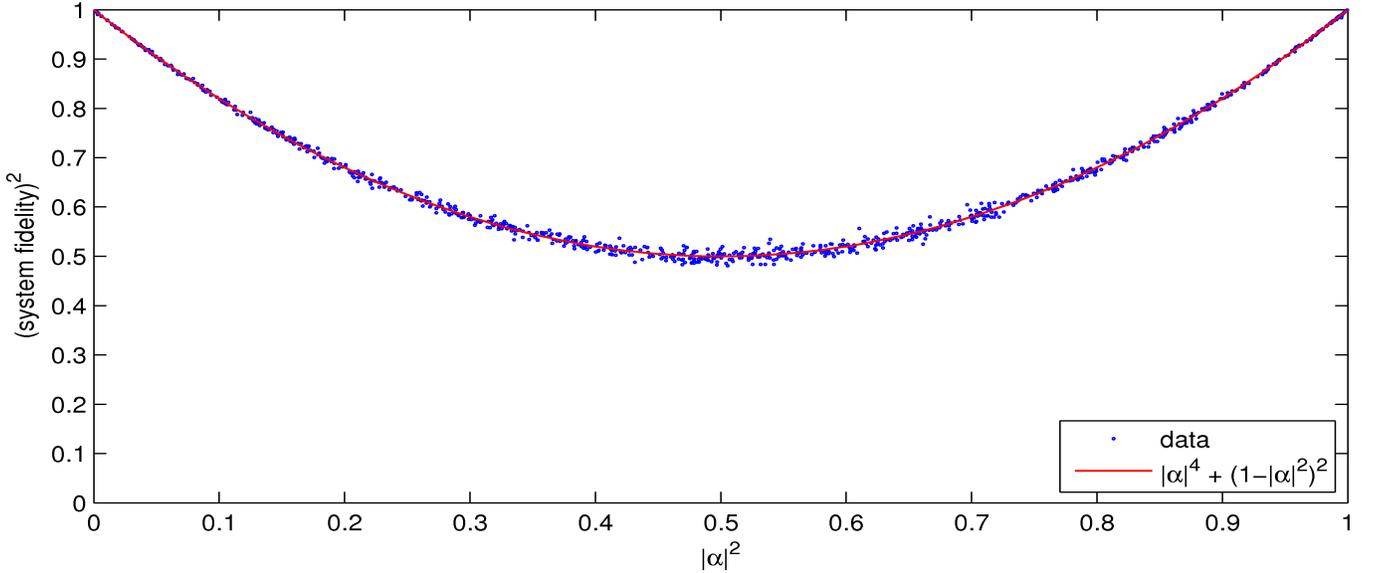
 
\begin{center}
\placefidplot{{fs_alpha}}
\caption{\label{fig:fs_alpha} 
The long term squared system fidelity, $\fid_s^2$, as a function of $|\alpha|^2$, where the initial system state is $\ket{\psi^s} = \alpha\ket{+_L} + \beta\ket{-_L}$ and $|\alpha|^2 + |\beta|^2=1$. The curve $y=|\alpha|^4 + (1-|\alpha|^2)^2$ is also shown, and the good fit is apparent. Each data point represents a random choice for $\alpha$ and $\beta$, as well as $V$, $\Henv$, and the initial environment state $\ket{\psi^e}$. The computational Hamiltonian is $\Hsys=X_L$ and $\Ep=128$.
Each data point is the average $\fid_s^2(\tfin)$ over the times  $\tfin = \lbrace 1,2,\ldots,10\rbrace \times 10^8$ to account for fluctuations in time of $\fid_s$ about the long term system fidelity.
}
\end{center}
\end{figure}

We next address the question of what the system fidelity falls to at late times for large $\Ep$.
For the Hamiltonian $\Hsys=X_L$, given $\ket{\psi^s}$ we can actually predict the long term system fidelity. 
To help uncover this relationship, we plot in Fig.~\ref{fig:XL_0L} the system fidelity as a function of time, with $\ket{\psi^s}$ taken to be $\ket{0_L}$. Note that the long term system fidelity is very near 1/2 for $\Ep\geq 16$. In Fig.~\ref{fig:XL_eig} we show the same thing but with $\ket{\psi^s}=\ket{+_L}$, an eigenstate of $X_L$, and see that the long term system fidelity is very near 1 for $\Ep \geq 16$. More generally, we observe that if we write
\begin{equation}\label{eq:psi_s_initial}
	\ket{\psi^s} = \alpha \ket{+_L} + \beta\ket{-_L} \,,
\end{equation}
with $\ket{\pm_L}$ being the codespace eigenstates of $X_L$ and $|\alpha|^2 + |\beta|^2 =1$, then the long term system fidelity is well approximated by $|\alpha|^4 + |\beta|^4$.
In Fig.~\ref{fig:fs_alpha} we show the long term system fidelity versus $|\alpha|^4 + (1-|\alpha|^2)^2$ for a set of randomly chosen $\ket{\psi^s}$ and the good fit is apparent. 

%
%

\begin{figure}
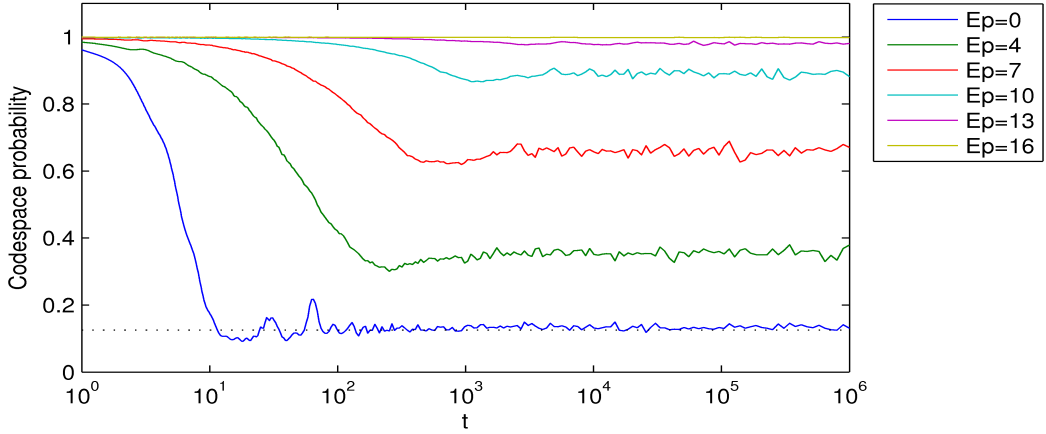
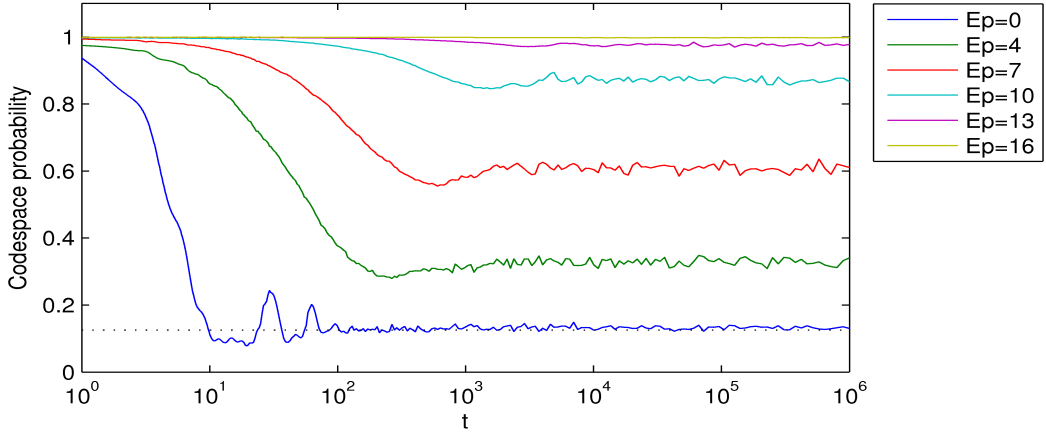
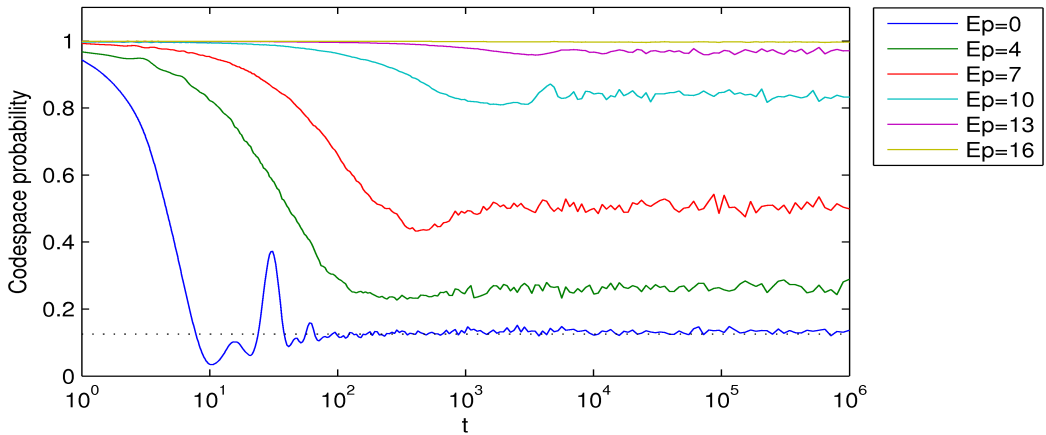
 
\begin{center}
	\subfigure[$\ket{\psi^s}=\alpha\ket{+_L}+\beta\ket{-_L}$]{\placeprobplot{{Pcodespace_rand}}} \\
    \subfigure[$\ket{\psi^s}=\ket{0_L}$]{\placeprobplot{{Pcodespace_0L}}}\\
    \subfigure[$\ket{\psi^s}=\ket{+_L}$]{\placeprobplot{{Pcodespace_eig}}}
\caption{ 
The probability $\bra{\phi(t)} P \ket{\phi(t)}$ of the system being found in the codespace for the
three different initial states used in Figs.~\ref{fig:XL_rand}, ~\ref{fig:XL_0L}, and ~\ref{fig:XL_eig}. The dashed line at 1/8 represents the expected probability for a maximally mixed 4-qubit state to be found in the 1-qubit codespace. In all three cases, for $\Ep \geq 16$ the codespace probability is very near 1 for all displayed times.
}
\label{fig:Pcodespace}
\end{center}
\end{figure}
We show in Fig.~\ref{fig:Pcodespace}, for the three choices of $\ket{\psi^s}$ displayed in Figs.~\ref{fig:XL_rand}, ~\ref{fig:XL_0L}, and ~\ref{fig:XL_eig}, the probability to remain in the codespace, $\bra{\phi(t)} P \ket{\phi(t)}$. We see that it is close to 1 for all displayed times for $\Ep \geq 16$, indicating that any loss of system fidelity is occurring because of errors inside the codespace. With $\Hsys = X_L$, the desired evolution, starting with the state in \eqn{eq:psi_s_initial}, is
\[
	\ket{\phi_0^s(t)} = \alpha e^{-it} \ket{+_L} + \beta e^{it}\ket{-_L}
\]
since the codespace eigenvalues of $X_L$ are $\pm 1$. 
Imagine that the only effect of the coupling to the environment is to induce dephasing in the $\Hsys$ energy eigenbasis. Then the density matrix of the system will approach
\newcommand{\bigbra}[1]{{\big\langle{#1}\big\vert}} 
\newcommand{\bigket}[1]{{\big\vert{#1}\big\rangle}} 
\[
	\rho(t) = |\alpha|^2 \bigket{+_L}\bigbra{+_L} + |\beta|^2 \bigket{-_L}\bigbra{-_L}
\]
and the squared system fidelity, $\bra{\phi_0^s(t)} \rho(t)  \ket{\phi_0^s(t)}$, is $|\alpha|^4 + |\beta|^4$.
That the data in Fig.~\ref{fig:fs_alpha} matches this is good evidence that the effect of the coupling to the environment is to cause dephasing in the energy basis of $\Hsys$.  

%
%
\newcommand{\toystatea}{{+}}
\newcommand{\toystateb}{{-}}

In our simulation we see that, for sufficiently large energy penalties, the system remains in the codespace and decoheres inside the codespace via dephasing of the energy eigenstates. We now present a simple phenomenological model that allows us to estimate $\tprot$, the time at which the effects of decoherence become appreciable. The model has three states. The first two states are the codespace eigenstates, $\ket{\toystatea}$ and $\ket{\toystateb}$, of the logically-encoded two-level computational Hamiltonian with energies $\omega$ and $-\omega$. The third state is a penalty state, representing all the states orthogonal to the codespace, and accordingly has energy $\Ep \gg \omega$. The third state is coupled to the first two as a result of interactions with the environment, so that the effective Hamiltonian is
\[
	H_\text{eff} = 
		\begin{bmatrix}
    	\omega & 0 & \lambda_\toystatea\\
    	0 & -\omega & \lambda_\toystateb \\
    	\lambda_\toystatea & \lambda_\toystateb & \Ep \\
  		\end{bmatrix}
  		\,.
\]
Here, $\lambda_\toystatea$ and $\lambda_\toystateb$ are the effective couplings of the first two states to the penalty state, and we assume that they are small compared to $\omega$. We imagine that $\lambda_\toystatea$ and $\lambda_\toystateb$ are proportional to some constant $\lambda$ that represents the overall scale of the effective couplings.
Expanding to lowest order in $\lambda_\toystatea$, $\lambda_\toystateb$, and $1/\Ep$, we find that 
\[
	\norm{\bra{\toystateb} e^{-i H_\text{eff} t} \ket{\toystatea}}^2 
	\lesssim \Big(\frac{\lambda_\toystatea\lambda_\toystateb}{\omega\Ep}\Big)^2 \,,
\]
so in this model the transition probability between states $\ket{\toystatea}$ and $\ket{\toystateb}$ is negligible for all time.

Treating the coupling as a perturbation, the effect of the coupling of the system states to the penalty state is to shift their energies.
The perturbed energies are calculated to be 
$$ E_{\pm} = \pm\omega - \frac{\lambda_{\pm}^2}{\Ep} $$ 
to lowest order in $\lambda_\toystatea, \lambda_\toystateb$, and $1/\Ep$. Thus in this little model, at time $t$, the interaction-induced phase difference between $\ket+$ and $\ket-$ is
\[
	\big(E_{\toystatea} - E_{\toystateb} - 2\omega  \big) t 
	= - \frac{\lambda_\toystatea^2 - \lambda_\toystateb^2}{\Ep} \,  t 
\]
so that the characteristic dephasing time is proportional to $\Ep / \lambda^2 \,.$
Generalizing from the toy model to an encoded two-level logical system with a coupling to the environment of size $\lambda$ and energy penalty term of size $\Ep$, we guess that
\[
	\tprot \propto \frac{\Ep}{\lambda^2} \,,
\]
in agreement with the behavior seen in Fig.~\ref{fig:protection_time}.

%
%
Returning to the simulation results, 
we have seen that a sufficiently large energy penalty keeps the system in the codespace, even for large $t$. 
We also presented evidence that decoherence inside the codespace occurs via dephasing in the energy basis. 
In particular, with a time-independent $\Hsys=X_L$, starting in an energy eigenstate, say $\ket{+_L}$, we find that for sufficiently large $\Ep$ the system remains approximately in that eigenstate for the duration of the simulation.
In adiabatic quantum computation \cite{FGGS00}, the state of the system is initially the ground state of a time-dependent computational Hamiltonian and, provided that the computational Hamiltonian is changed slowly enough, the evolving state is expected to remain near the instantaneous ground state. One might therefore expect good fidelity in the adiabatic computation case as well.

We now show the results of simulations for the one logical qubit adiabatic computation 
\[
	\Hsys(t) = \left(1-\frac{t}{\tfin}\right) X_L + \frac{t}{\tfin} Z_L \,,
\]
where the initial system state $\ket{\psi^s} = \frac{1}{\sqrt2} (\ket{0_L} - \ket{1_L})$ is the ground state of $\Hsys(0)$. 
The results are shown in Fig.~\ref{fig:adiabatic} for $\tfin=10,000$.
Observe that for $\Ep \geq 16$, the system fidelity remains very high for the duration of the computation.

\begin{figure}
\begin{center}
\placefidplot{{fs_t_adiabatic}}
\placefidplot{{f_t_adiabatic}}
\caption{\label{fig:adiabatic} 
For the adiabatic computation, $\Hsys(t) = (1-\frac{t}{\tfin}) X_L + \frac{t}{\tfin} Z_L$ with $T=10000$, the
(top) squared system fidelity, $\fid_s^2$, and (bottom) squared total fidelity, $\fid^2$,
as functions of time $t$ for $\lambda=0.1$.
All data are for a particular random instance of $\Henv, V$, and $\ket{\psi^e}$ with $\nenvactual$ environment qubits, with the system initially in the ground state of $\Hsys(0)$. Note that for $\Ep \geq 16$, we have nearly perfect system fidelity throughout the evolution.}
\end{center}
\end{figure}

%
%
We emphasize that our numerical results are for a small system (1 logical qubit made of 4 physical qubits) coupled to a small environment (of $\nenvactual$ qubits). 
We do not know if the observations we have made for one logical qubit will hold in more complicated systems with many logical qubits. In particular, we would like to know if with a large number of qubits, modest energy penalties can keep the system in the codespace and, if inside the codespace, whether the decoherence is limited to dephasing in the energy basis. If so, this would be of great help in protecting adiabatic quantum computation. 
Furthermore, we are concerned that in our simulations, the size of the environment may be too small, especially given the large values of $\Ep$ that we are exploring. It would be disappointing if our encouraging small system simulation results are artifacts of having too small an environment or do not reflect what actually happens in large systems.
Nevertheless, these numerical results, in conjunction with the proof that the energy penalty method works in the infinite $\Ep$ limit, suggest that the energy penalty method may be a useful approach towards the development of fault-tolerant Hamiltonian based quantum computing.

%
\section{Conclusion}
%

In this paper, we considered the energy penalty method of error suppression, i.e., the method of achieving error suppression by encoding a Hamiltonian using a quantum error detecting code and adding a constant term that penalizes states outside of the codespace. We proved that this method does indeed work in principle. Specifically, we showed that, in the limit of an infinitely large energy penalty, the actual evolution of the system is precisely the evolution in the absence of unwanted control errors and environmental interactions, provided that the code can detect these errors. Moreover, we have provided some bounds governing the finite energy penalty scenario, allowing one to bound the energy penalty required to attain the desired evolution with good fidelity. 
We believe that these bounds can be improved, as supported by our numerical evidence for a single logical qubit, and leave their tightening as an interesting open problem. We hope that progress in this area will eventually lead to a practical fault-tolerant paradigm for Hamiltonian based quantum computation.

%
%
\section*{Acknowledgments}
The authors are grateful to Robin Blume-Kohout, Daniel Lidar, Seth Lloyd, and Peter Shor for valuable discussions. 
E.F. thanks members of the Google Quantum Artificial Intelligence Lab for their input.
We gratefully acknowledge support from 
the US Department of Energy under cooperative research agreement contract no. DE-FG02-05ER41360, 
the Army Research Office under grant no. W911NF-12-1-0486, 
and from the National Science Foundation under award no. CCF-121-8176.
L.Z. acknowledges support from the MIT Undergraduate Research Opportunities Program.

%
%
\section*{Appendix} \label{sec:appendix}
In this paper we focused on the simplest case, where $V$ acts 1-locally on the system and the quantum error detecting code can detect 1-qubit errors. In this appendix we show that this simplification is not necessary. As long as the error detecting code can detect the errors that $V$ causes, our infinite energy penalty theorem still holds. This includes, for example, the case where $V$ acts $k$-locally and the code can detect $k$-local errors. 
%
Specifically, the only requirement on $V$ is that 
\begin{equation}\label{eq:PVP}
	P V P = 0 \,.
\end{equation}
We now present a proof of this general case.

Define $R_r$ (for $r=0,\ldots,\nlogical$) to be 
\[
	R_r = \sum \big\lbrace A_1 \otimes \cdots \otimes A_\nlogical : A_i \in \lbrace \P_i, \Q_i \rbrace \text{ such that } |\lbrace i : A_i = Q_i \rbrace | = r \big\rbrace \,,
\] 
where as before, $\P_i$ is the codespace projector for the $i$th logical qubit and $\Q_i = \id - \P_i$. 
In other words, $R_r$ is 
the sum of all terms, each of which is a tensor product of a total of $\nlogical$ $\P_i$'s and $\Q_i$'s, one for each logical qubit, such that precisely $r$ of these projectors are $\Q_i$'s.
For example, $R_0 = P$, $R_\nlogical = Q_1 Q_2 \cdots Q_\nlogical$, and 
\[
	R_1 = Q_1 P_2 \cdots \P_\nlogical \ \  + \ \ \cdots \ \ + \ \ P_1\cdots P_{\nlogical-1} Q_{\nlogical} \,.
\] 
Observe that the $R_r$ are in fact a complete set of orthogonal projectors:
\[
\begin{split}
	R_r ^2 &= R_r \qquad\text{ for all } r \\
	R_r R_{r'} &= 0 \qquad\text{ for } r\neq r' \, \\
	\sum_{r=0}^\nlogical R_r &= \id \,,
\end{split}
\]
where the last equality can be obtained by expanding out $\id = \prod_i (\P_i + \Q_i)$.

Now, recall that $e^{i \Ep \tau Q_i} P_i = P_i$ and that $e^{i \Ep \tau Q_i} Q_i = e^{i \Ep \tau} Q_i $. Therefore, using the definition of $\tildeQ$ in \eqn{eq:tildeQ}, we see that for any $r$,
\[
\begin{split}
	U_P^\dag(\tau) R_r 
	&= e^{i \Ep \tau \tildeQ} R_r \\
	&= \prod_{i=1}^\nlogical e^{i \Ep \tau Q_i} R_r \\
	&= e^{i r \Ep \tau} R_r	
\end{split}
\]
because each term in $R_r$ consists of precisely $r$ $Q_i$'s. 
Applying $U_P^\dag$ to $\id = \sum_{r=0}^\nlogical R_r$ therefore lets us write
\[
	U_P^\dag (\tau) = \sum_{r=0}^\nlogical e^{i r \Ep \tau} R_r \,
\]
so that applying $U_P^\dag$ to $V \P$ gives
\[
	U_P^\dag V P = \sum_{r=0}^\nlogical e^{i r \Ep \tau} R_r V P \,.
\]

We now apply our key requirement of \eqn{eq:PVP} to see that the $r=0$ term is $R_0 V P = PVP = 0$. Thus, we have
\[
	U_P^\dag V P = \sum_{r=1}^\nlogical e^{i r \Ep \tau} R_r V P \,,
\]
instead of the 1-local version in \eqn{eq:UpVP} that arose because $R_r V P = 0$ for $r \neq 1$ in that case. Our formula for $F$ from \eqn{eq:Fmodulated} therefore generalizes to
\[
	F(t) = \sum_{r=1}^\nlogical \int_0^t e^{i r\Ep \tau} R_r U_0^\dag(\tau) V(\tau) U_0(\tau) \P \dd\tau \,.
\]
Note that every term in $F(t)$ has a phase of $e^{i r \Ep \tau}$ for some $r>0$. Applying the Riemann-Lebesgue lemma, we again conclude that in the infinite $\Ep$ limit, $F(t) \rightarrow 0$ and our theorem follows.
This form of $F$ may be useful in deriving finite energy penalty bounds in the case that we have a code that can protect against more than 1-local errors.

%
%

\end{document}